\documentclass{aa} 
\usepackage{graphicx}
\usepackage{txfonts}
\usepackage{array}
\PassOptionsToPackage{hyphens}{url}\usepackage[breaklinks]{hyperref}
\usepackage{natbib}
\newcommand{\mygi}{MyGIsFOS}

\newcommand{\logg}{\ensuremath{\log\,g}}
\def\teff{$T\rm_{eff}$}
\newcommand{\kms}{$\rm km s ^{-1}$}

\newcommand{\ymp}{\hbox{TYC\,170--1218--1}}

\begin{document} 

\title{TYC 170--1218--1: A new  r-process-enhanced extremely metal-poor star, rich in Th.
\thanks{Based on observations made with UVES at VLT 112.25EH.001 and
with Mike at the Magellan Clay telescope.}
}
\titlerunning{TYC 170--1218--1}

\author{
E.~Caffau    \inst{1} \and
P.~Bonifacio \inst{1} \and
L.~Sbordone \inst{2} \and   
L.~Monaco \inst{3} \and
J.~Alazzawi \inst{1} \and
M.~Spite  \inst{1} \and
P.~Fran\c{c}ois \inst{4,5} \and
A.~Mucciarelli \inst{6}
}

\institute{LIRA, Observatoire de Paris, Universit{\'e} PSL, Sorbonne Universit{\'e}, Universit{\'e} Paris Cit{\'e}, CY Cergy Paris Universit{\'e}, CNRS,92190 Meudon, France
\and
European Southern Observatory, Casilla 19001, Santiago, Chile
\and
Universidad Andres Bello, Facultad de Ciencias Exactas, Departamento de F\'isica y Astronom\'ia – Instituto de Astrof\'isica, Autopista Concepci\'on-Talcahuano 7100, Talcahuano, Chile
\and
LIRA, Observatoire de Paris, Universit{\'e} PSL, Sorbonne Universit{\'e}, Universit{\'e} Paris Cité, CY Cergy Paris Universit{\'e}, CNRS,75014 Paris, France
\and
UPJV, Universit\'e de Picardie Jules Verne, 33 rue St Leu, 80080 Amiens, France
\and
Dipartimento di Fisica e Astronomia, Universit\`a degli Studi di Bologna, Via Gobetti 93/2, I-40129 Bologna, Italy
}

   \date{Received July 21, 2025; accepted August 13, 2025}

  \abstract
{Extremely metal-poor (EMP) stars are formed from  gas clouds enriched by one or a few supernova explosions belonging to the first stellar generation 
and this very limited number of sources of metal enrichment is suitable to produce peculiar chemical patterns that give birth to stars with anomalous chemical composition. 
Among the EMP stars, r-II stars are characterised by an over-abundance of the heavy elements with respect to iron.
}
{In the search for apparently young, metal-poor stars, 
we serendipitously selected \ymp\ , which 
turned out to be an EMP star, enhanced in neutron capture elements over iron.
Our aim is to obtain a detailed chemical inventory for this exceptional object.
}
{We investigated  high-resolution spectra observed with UVES at the VLT telescope and Mike
at the Magellan Clay telescope. 
We derived the abundance of 33 elements using the \mygi\ code and an ATLAS\,9 model atmosphere.
}
{The star is an EMP with [Fe/H]=--3.52. It is enhanced in the $\alpha$ elements, as EMP stars usually are. It is an r-II star with [Eu/Fe]=+1.84 and [Th/Fe]=+1.85.
The star is also poor in carbon with respect to iron. 
The quality of the spectra was insufficient for us to detect uranium. Kinematically the star belongs now to the Galactic halo, but it joined the Milky Way during the Sequoia accretion event.}
{}

   \maketitle
%
\nolinenumbers
\section{Introduction} \label{intro}
 Metal-poor (MP) stars are precious witnesses of the early chemical evolution
of our Galaxy. The class of extremely metal-poor (EMP) stars that includes
all the stars with $\rm -4 \le \mathrm [Fe/H] < -2.8 $ \citep{bonifacio2025},
has been polluted very little by the chemical elements produced in
supernovae, neutron star mergers, and any other event that contributes
to the chemical build-up of the early Galaxy.
The small number of polluters in the EMP regime is responsible for the large variety of 
chemical patterns in the heavy elements of EMP stars.

The elements lighter than iron are expected to be produced in the stellar interiors through  nuclear fusion
reactions, 
while the elements heavier than the iron peak are synthesised through neutron captures \citep[see e.g.][]{sneden2008,arcones2023}.
The neutron captures can occur at different rates depending on the neutron
density: {\em i)} slow, $s$ process, that is the time between two neutron captures
is longer than the time for $\beta$ decay of the produced nucleus,
for neutron densities of $n \approx 10^5$ cm$^{-3}$; {\em ii)} intermediate, $i$ process
with neutron densities in the range of $10^{14}-10^{16}$cm$^{-3}$;
{\em iii)} rapid, $r$ process with neutron densities of $> 10^{23}$ cm$^{-3}$ \citep{cowan1977}. 
In both the $i-$ and $r-$process, several neutrons can be captured
before the nucleus produced by the previous capture has time to $\beta$ decay.
The sites of the $r$ process are widely debated \citep{arcones2023}, the main candidates
being neutron star mergers, supernovae, neutrino-driven winds, and jets
in highly rotating supernovae. Recently \citet{wanajo2024} 
proposed neutron star-black hole mergers as a viable source of
both solar-like $r-$process patterns and actinide boost patterns, the latter
for a large enough electron fraction (proton number per nucleon), $Y_e>0.05$, 
in the dynamical ejecta of the merger.

Europium is an almost pure r-process element ($\sim 95$\,\% is from the r-process; \citet{sneden2008}) and the fact that in the MP regime [Eu/Fe] spans almost 4\,dex
means that substantially diverse chemical enrichments through the r process characterised the young Universe.
Based on the europium abundance, \citet{christlieb2004} classified
MP stars as r-I ($\rm 0.3<[Eu/Fe]<1.0$ and $\rm [Ba/Eu]<0$) and r-II ($\rm [Eu/Fe]>1.0$ and $\rm [Ba/Eu]<0$).
The number of EMP r-II stars, according to this classification, is limited, but they are not really rare objects \citep[several tens in the SAGA database,][are available]{SAGA}.
Another possible classification is provided by \citet{holmbeck2020} for r-I ($\rm 0.3<[Eu/Fe]\leq 1.0$ and $\rm [Ba/Eu]<0$) and r-II ($\rm [Eu/Fe]>0.7$ and $\rm [Ba/Eu]<0$) stars,
but we find it safer, in the discussion of r-II stars, to keep the classification by \citet{christlieb2004}.

The abundances of the radioactive actinide elements (with an atomic number in the range of 89-102) are difficult to derive from the stellar spectra, especially 
in the MP and EMP regimes, because they have a low abundance, and their absorption lines in the spectrum are often 
too weak to be detected.
Just a few EMP stars with a Th abundance determination are known; the majority of them show $\rm [Th/Fe]>1$ \citep[see e.g.][]{placco2023} and
a subsample of these stars also have a U abundance determination.
Clearly an enhancement in Th and U makes it possible to derive their abundances in the EMP regime \citep[see e.g.][]{hill2002}.
The mechanism that produces this enhancement can be the same as the one that makes the star an r-II, but there can be an extra channel (a boost) to enrich the actinides further than with the r process.
Th-rich stars are the only MP stars among which it has been possible to measure the abundance of uranium and for which an age estimate was made using the U/Th ratio \citep[see e.g.][]{ludwig2010}. 
Since the U/Th production ratio is much less sensitive to the physical conditions
under which the r process takes place than the Th/Eu production ratio,
it is  important, for any star with a Th enhancement, to have a spectrum quality, both in terms
of spectral resolution and the signal-to-noise ratio (S/N),  sufficient to derive the U abundance. We here investigate  high-spectral-resolution spectra of the star
\ymp, which is strongly enhanced in all neutron
capture elements (making it an r-II star) and also strongly enhanced in  Th.

\section{Observations and analysis} 

\begin{figure}[hb]
    \centering
   \resizebox{0.48\textwidth}{!}{\includegraphics{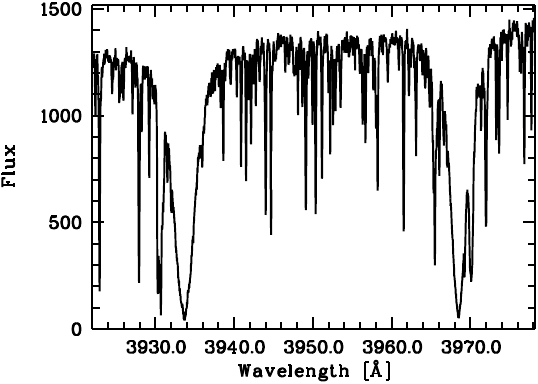}}
   \caption{UVES observed spectrum in the wavelength range of the \ion{Ca}{ii}-H and -K lines.} 
    \label{fig:cahk}
\end{figure}

We analysed one 3000\,s   UVES \citep{dekker2000} spectrum of \ymp\ 
observed in the ESO programme 112.25EH.001 (P.I. P. Bonifacio) on the night of December 4 2023.
The star was observed  in the setting DIC2 437+760, with a slit of 0\farcs{5}  in both  arms (corresponding to a resolving power of $\sim$70,000).  
The spectrum does not show any sign of stellar activity (the profile of H$\alpha$ is regular and no absorption is visible in the core of the \ion{Ca}{ii}-H and -K lines as shown in  Fig.\,\ref{fig:cahk}). The S/N per pixel is 80 at 400\,nm and 100 at 650\,nm.
The star was targeted because it has a high transverse velocity with respect to the Sun and is in a region of the colour-magnitude
diagram that is also populated by  MP stars that are apparently young \citep[see][figure 2]{bonifacio2024}.
Thus the finding that it is an r-II star was serendipitous.
The star is contained in the \textit{Gaia}\,DR3 release \citep[Gaia\,DR3\,3116478233336506624;][]{gaiadr3} 
and we used the astrometric and photometric data available.

Recently, we secured three 1200s exposures
 of \ymp\ with Mike at the Magellan Clay 6.5\,m telescope on January 31 2026 (P.I. L. Monaco). 
We used a slit of 0\farcs{7} and
$2\times 2$ on-chip binning in both the red and blue arms. 
This provides a resolving power of 41\,500 in the blue arm and 32\,500\ in the red arm,
the bluest orders of the blue arm are slightly undersampled.   
The blue arm covers the range 334\,nm to 499\,nm and the red arm 
the range 484\,nm to 915\,nm. 
The three exposures were co-added and 
the S/N per pixel on the co-added spectrum is 153 at 400\,nm and 
250 at 650\,nm.  
On  March 10 2026, we acquired another spectrum
of 1200\,s exposure, 
with Mike, with the same set-up except that the binning was $1\times 1$.
Both the UVES and the two Mike spectra were used to determine the star's radial
velocity by template matching using a dedicated synthetic spectrum.

\subsection{Kinematics} \label{sec:kine}

The radial velocity of \ymp\ available in \textit{Gaia}\,DR3 is
$257.46\pm  0.37$\,\kms. 
The UVES spectrum provides 256.7\,\kms, measured from the
spectral range around the IR \ion{Ca}{ii} triplet, and is therefore similar to the 
range of the \textit{Gaia} RVS. A model of the terrestrial atmosphere \citep{tapas2025}
was used to provide a zero point to the measured radial velocity. 
In the same spectral region and with correction of the zero point, 
the Mike spectrum  of January 31 provides a radial velocity of 256.3\,\kms\ ,
while the one of March 10 provides 256.9\,\kms.
For both the UVES and the Mike spectrum, we assume an  error of 0.5\,\kms\
dominated by centring the star on the slit, in spite of our zero point correction for the telluric lines.
The four radial velocities (an average Gaia value, one UVES and two Mike) refer to four different epochs 
(2016 Gaia, 2023 UVES, January and March 2026 Mike)
and they show consistency, within the uncertainties.

In order to investigate the star's kinematics, 
we used the {\tt galpy} code \citep{bovy15}, along with its standard potential MWPotential2014 and the \textit{Gaia}\,DR3 astrometric parameters and radial velocity. The parallax was zero-point corrected according to the prescriptions of \citet{zp}.  

\begin{figure}[hb]
    \centering
   \resizebox{0.48\textwidth}{!}{\includegraphics{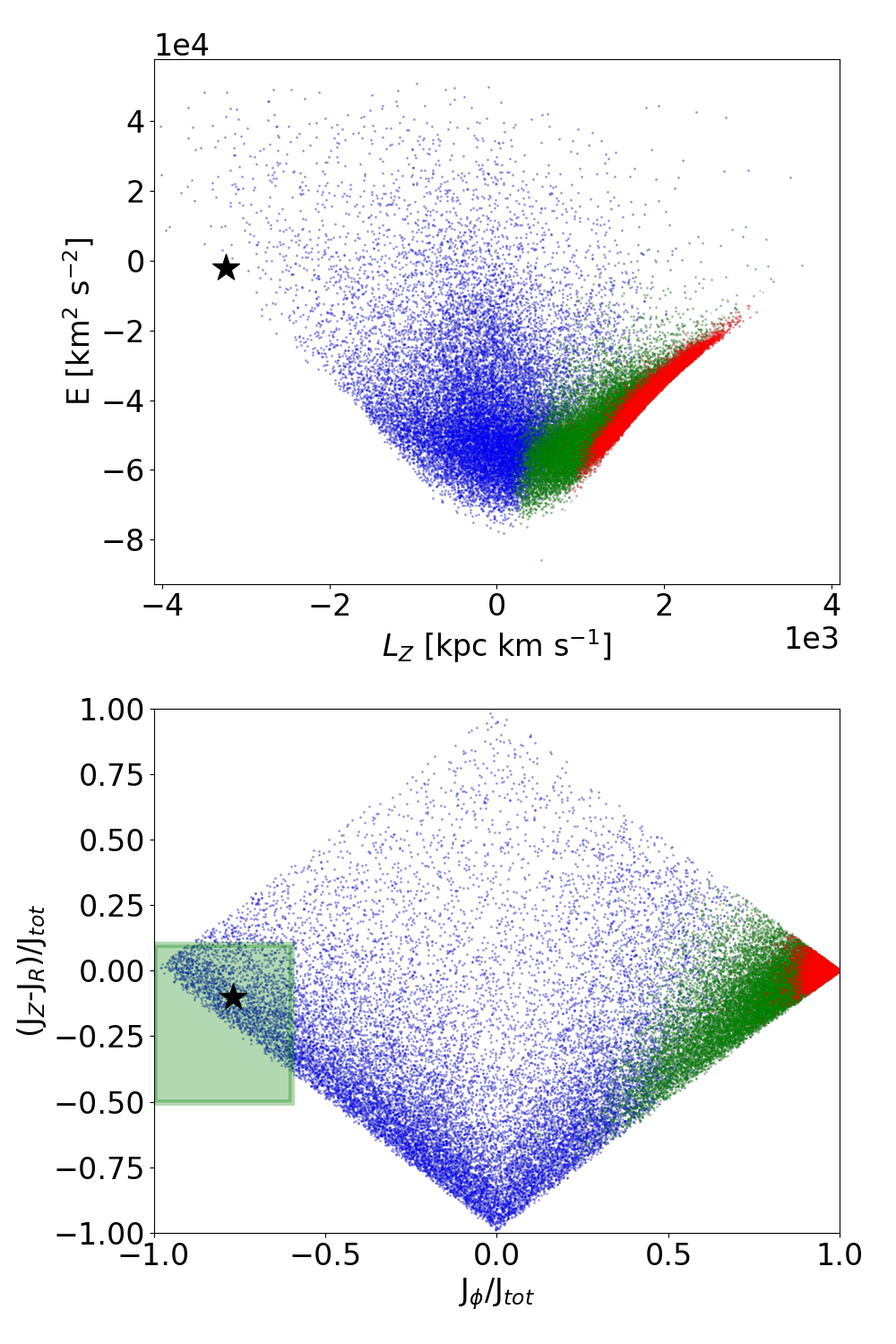}}
\caption{Position of \ymp, (filled black star) in the energy, E, versus angular momentum, L$_Z$ (top panel), plane and in the action diamond plane, namely the difference in the vertical and radial actions ($J_Z-J_R$) versus the azimuthal action, $J_\phi$ (equal to the vertical component of the angular momentum, $L_Z$), all normalised to the total action, defined as $J_{tot}=|J_\phi|+J_R+|J_Z|$. Stars from the 'good-parallax' sample of \citet{bonifacio21} are also shown for comparison and divided into halo (blue), thin disc (red), and thick disc (green) stars. The shaded green rectangle marks the region where likely Sequoia members are expected to be found. 
    }
    \label{fig:kine}
\end{figure}

\begin{figure}[hb]
    \centering
   \resizebox{0.48\textwidth}{!}{\includegraphics{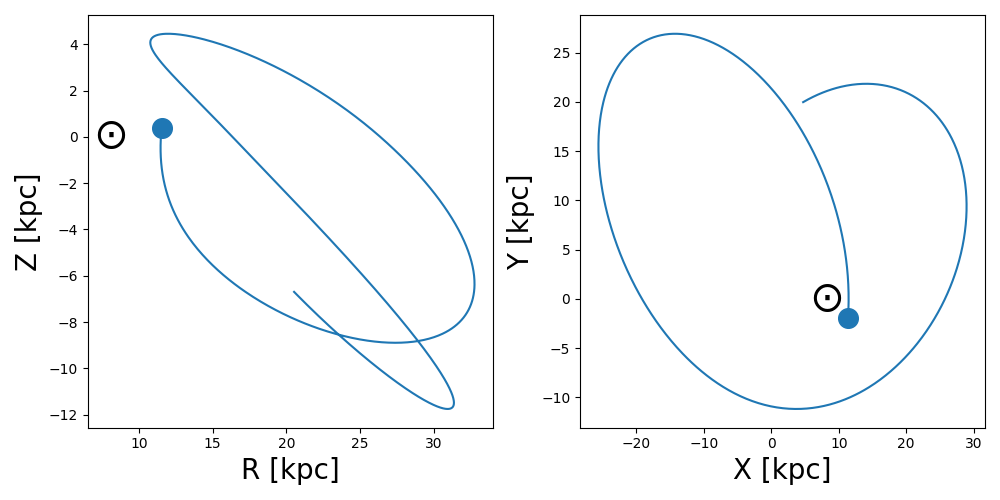}}
   \caption{Orbit of \ymp, integrated for one gigayear in the past. The positions of our target (filled blue circle) and the Sun (black solar symbol) are marked.}
    \label{fig:orbit}
\end{figure}

Figure\,\ref{fig:kine} presents the position of \ymp\, (black star) in two widely used kinematic planes: (upper panel) the orbital energy versus the angular momentum and (lower panel) the action diamond. The background population is the `good-parallax' sample of \citet{bonifacio21}, where stars are identified as belonging to the halo (blue), the thin disc (red), and the thick disc (green), according to the \citet{bensby14} criteria.
The shaded green rectangle marks the region where likely Sequoia \citep{barba2019} members are expected to be found according to the criteria defined in \citet{feuillet21}.
As can be seen in Fig.\,\ref{fig:kine}, \ymp\ belongs to the Galactic halo and more specifically to the Sequoia accretion event. 
In Fig.\,\ref{fig:orbit} the orbit of the star, integrated for the past gigayear, is shown. In the figure, X, Y, and Z are Galactocentric Cartesian co-ordinates, while R is the cylindrical radius, $R=\sqrt{X^2+Y^2}$. The current position of \ymp\, is marked by the filled blue circle, while the position of the Sun is also shown in black.

\subsection{Stellar parameters and chemical analysis}

We derived the stellar parameters by using \textit{Gaia}\,DR3 photometry and the zero-point corrected parallax, as done by \citet{caffau2024}.
The reddening ($\rm A_0 =0.68$) is from the maps by \citet{vergely2022}. The dereddened colour $ (G_{BP} - G_{RP})_0$, compared to synthetic colours
from the KOALA database \citep{KOALA}, provided the effective temperature (\teff).
The surface gravity (\logg) was derived by using the Stefan-Boltzmann law with the distance  inferred by the zero-point corrected parallax.  
The uncertainty in the \textit{Gaia}\,DR3 photometry is so small (0.003\,mag in G and 0.005\,mag in $(G_{BP} - G_{RP})$) to be non-influent in deriving the stellar parameters.
The \textit{Gaia}\,DR3 parallax has a relatively small uncertainty ($\rm 0.229\pm 0.017$) that translates into an uncertainty in \logg\ within 0.06\,dex.
This is consistent with the perfect balance of the iron abundance, A(Fe), derived from ionised and neutral Fe lines, once the latter have been  corrected for departure from local thermodynamical equilibrium (NLTE).
In the case in which the zero-point on the parallax is not applied, the \logg\ is 0.09\,dex smaller.
By using the calibration by \citet{mucciarelli2021}, the \logg\ is 0.02\,dex larger and the \teff\ 48\,K hotter.
We then assume an uncertainty in \teff\ of 100\,K, twice the difference of \teff\ from the two methods and 0.1\,dex in \logg.  

We adopted a first guess metallicity of --3 to derive the initial stellar parameters. We then updated the metallicity with the value obtained by \mygi\ \citep{sbordone2014} in the chemical analysis to update the parameters.
A detailed description of our iterative procedure can be found in \citet{lombardo2021}.
We initially derived the parameters assuming a mass of 0.8\,$M_\odot$, then we used these parameters to
determine the age and the mass with the Bayesian inference code SPInS \citep{LebretonReese2020,spins2020}, 
which provided an estimate of the mass as 0.87\,$M_\odot$. 
We used SPInS with a flat prior on the age of between 0 and 13.8\,Gyr, to avoid unphysical ages larger
than the age of the Universe. We used the BaSTI evolutionary tracks enhanced by 0.4\,dex in the
$\alpha$ elements \citep{pietrinferni2021}. The input parameters were log(\teff), $\log (L/L_\odot)$, and
[Fe/H]. The result is $\left(10\pm 2\right)$\,Gyr.
We then re-ran the procedure with the derived mass.
The procedure in deriving the stellar parameters and the mass was iterated.
The final stellar parameters (see Table\,\ref{tab:param}) are: \teff\ of 4581\,K, \logg\ of 1.33, and microturbulence of 2.0\,\kms, from 
the calibration of \citet{mashonkina2017}.

In a first step of the abundance analysis, we used \mygi\ in the standard `multi-model' mode \citep{sbordone2014}
using  synthetic grids
computed with SYNTHE \citep{kurucz} using our grid of ATLAS\,12 models \citep{kurucz} covering metallicities from --4.0 to --0.5 in steps of 0.5\,dex,
effective temperatures from 4000\,K to 5200\,K in steps of 200\,K, surface gravities from 0.5 to 3.0 in steps of 0.5\,dex, and $\alpha$ enhancement from --0.4 to +0.4\ in steps of 0.4\,dex.
Conscious that the star has a non-solar-scaled abundance pattern,
we switched to the `single model' use of \mygi\ as described in \citet{caffau2024}. 
With the final parameters, we computed with Turbospectrum \citep{alvarez1998,plez2025} a grid of synthetic spectra (with a single \teff\ and \logg, and three values of microturbulence of 1, 2, and 3\,\kms) by using an ATLAS\,9 model 
that we computed using the parameters (\teff\ and \logg) selected for the star, 1\,\kms microturbulence, [$\alpha$/Fe]=+0.4, a metallicity of $-3.50$, and the opacity distribution function
from the KOALA  database.
In the grid the pivot abundance value for each element, X, was not the solar scaled value but the A(X) value derived from the first `multi-model' run of \mygi.
The process was iterated, changing the pivot values (and recomputing the grid of synthetic spectra) up to a difference in the derived abundance and the pivot value inferior to the grid abundance step (0.2\,dex). The grid was finally recomputed by putting as a pivot for carbon A(C) the value derived from the fit of the G-band (see below).
The derived abundances are provided in Table\,\ref{tab:abbo}.
The uncertainty ($\sigma$) is the line-to-line scatter. For the elements whose abundance is based on a single line, we attributed an uncertainty of 0.15\,dex.
In the table we provide the NLTE correction and the reference for it.
The lines used are provided in a table in electronic form at the CDS.
As a sanity check, we ran \mygi\ on the Mike spectrum of January 31, and the results were 
compatible with those of UVES, within less than 1\,$\sigma$ for any
abundance. We prefer not to combine the two spectra given the different
sampling and resolution. We did not run \mygi\ on the Mike spectrum of 
March 10 that has a lower S/N, due to the shorter exposure time. 

\begin{table*}
\caption{Parameters  of \ymp.}
\begin{tabular}{rlrlllll}
\hline
  \multicolumn{1}{c}{\textit{Gaia} DR3 ID} &
  \multicolumn{1}{c}{NAME} &
  \multicolumn{1}{c}{Teff} &
  \multicolumn{1}{c}{log g} &
  \multicolumn{1}{c}{VTURB} &
  \multicolumn{1}{c}{[Fe/H]} &
  \multicolumn{1}{c}{log$(L/L_\odot)$} &
  \multicolumn{1}{c}{$\sigma_L$} \\
   & & K & [c.g.s] & \kms &  dex & & \\
\hline
  3116478233336506624 & \ymp\ & 4579 & 1.35 & 2.01 & --3.52 & 2.63 & 0.06\\
\hline
\label{tab:param}
\end{tabular}
\end{table*}

\begin{figure}[hb]
    \centering
   \resizebox{0.48\textwidth}{!}{\includegraphics{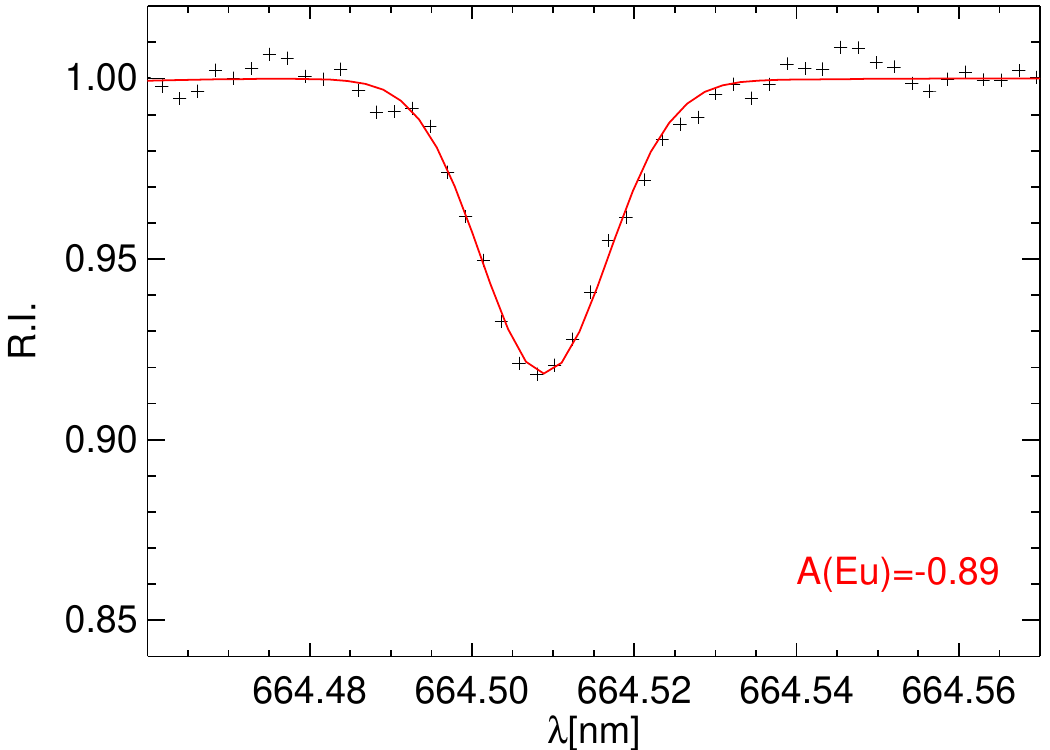}}
   \resizebox{0.48\textwidth}{!}{\includegraphics{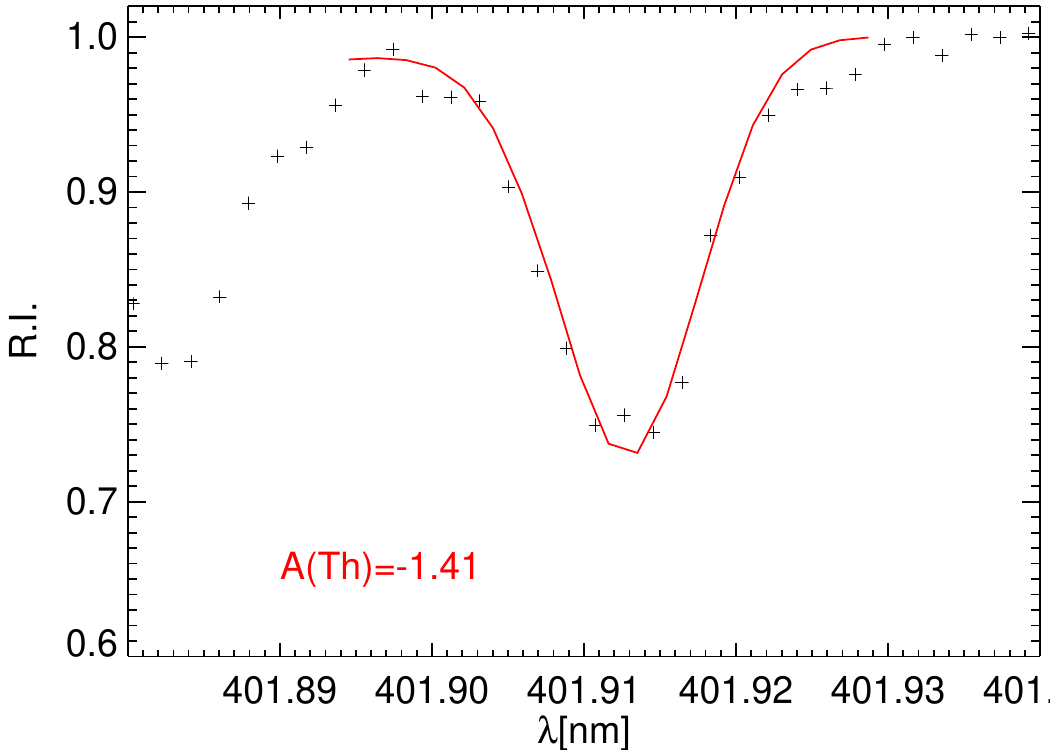}}
   \caption{UVES spectrum (black crosses) compared to the best fit profile (solid red) in the region of: (upper panel) the 664.5\,nm \ion{Eu}{ii} line and (lower panel) the 401.9\,nm \ion{Th}{ii} line. 
    }
    \label{fig:euth}
\end{figure}

The star \ymp\ is an EMP, enhanced in $\alpha$ elements, as expected, and it is rich in n-capture elements. 
What is noticeable is the abundance of Eu and Th that are both strongly enhanced
([Eu/Fe]=+1.84 and [Th/Fe]=+1.85\ in LTE with Fe from ionised lines).
In Fig.\,\ref{fig:euth} one line for each of these two elements is shown.
In fact, all elements belonging to the second (Ba, La, Ce, Pr, Nd, Sm, Eu, Gd, Dy, and Er) or third (Th) peak have $\rm [X/Fe]>1$ (see Table\,\ref{tab:abbo}), while the elements of the first peak (Sr, Y, Zr) are just slightly enhanced ($\rm 0<[X/Fe]<1$).

\begin{figure}[hb]
    \centering
   \resizebox{0.48\textwidth}{!}{\includegraphics{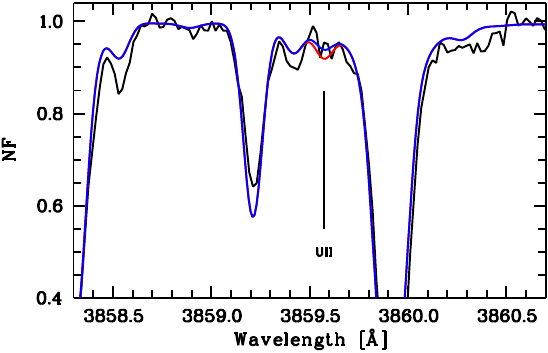}}
   \caption{UVES spectrum (solid black) in the region of the \ion{U}{ii} line at 385.9571\,nm compared to synthetic spectra with A(U)=--1.8 (solid red) and A(U)=--2.0 (solid blue). 
    }
    \label{fig:u}
\end{figure}

Unfortunately, the S/N of our UVES spectrum is not sufficient to detect the \ion{U}{ii} 385.9\,nm resonance
line. 
The Mike spectrum in this wavelength region has a slightly lower S/N than the UVES spectrum.
Its  lower resolution and undersampling prevent the situation from significantly improving
with respect to the UVES spectrum.
Even the upper limit we provide for U is not significative (see Fig.\,\ref{fig:u}), since it is higher than the measured [Th/Fe], while in all
stars in which U has been measured [U/Fe]$\le$[Th/Fe]. Further observations are needed to elucidate this point.

\begin{figure}[hb]
    \centering
   \resizebox{0.48\textwidth}{!}{\includegraphics{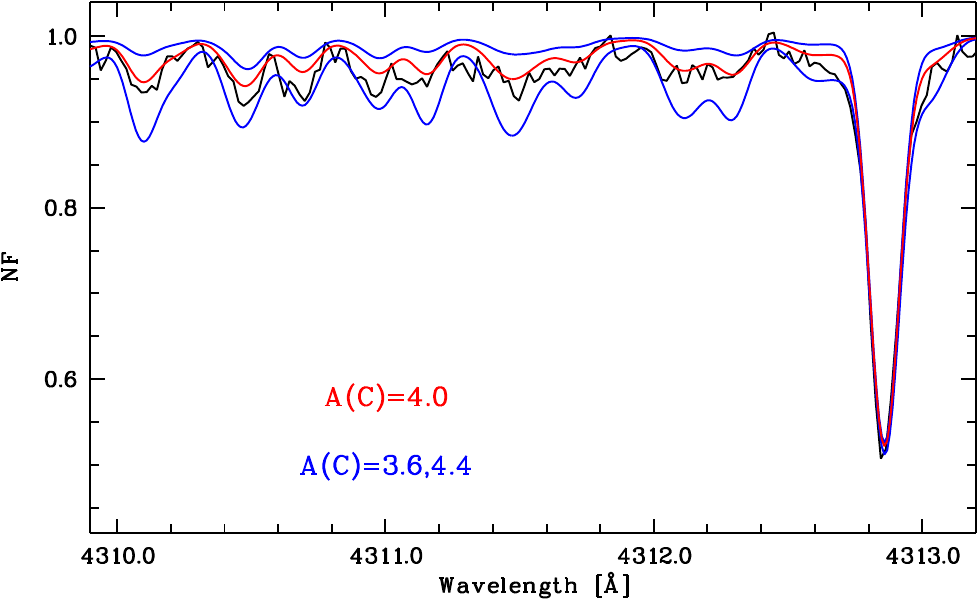}}
   \caption{UVES spectrum (solid black) in the region of the G-band compared to synthetic spectra. 
    }
    \label{fig:gband}
\end{figure}

Carbon was derived by line profile fitting on the G-band (see Fig.\,\ref{fig:gband}) and provided A(C)=4.02, which is lower than expected due 
to the mixing related to the stellar evolution.
In fact, according to \citet{placco2014}, a depletion of about 0.5\,dex is expected due to the evolution of this star, while the star has a ratio [C/Fe] of almost --1\,dex.
We were not able to get any insight into the $\rm ^{12}C/^{13}C$ ratio because the G-band is too weak.

\begin{table*}
\caption{Abundances of \ymp. For [X/Fe], the iron abundance [Fe/H] is from neutral lines when the abundance A(X) is based on neutral lines, from \ion{Fe}{ii} lines when A(X) is from ionised lines.}
  \label{tab:abbo}
\begin{tabular}{lrrrrrrrrrl}
\hline
  \multicolumn{1}{l}{Element} &
  \multicolumn{1}{r}{Z} &
  \multicolumn{1}{r}{ion} &
  \multicolumn{1}{r}{N lines} &
  \multicolumn{1}{r}{$\rm A(X)_\odot$} &
  \multicolumn{1}{r}{A(X)} &
  \multicolumn{1}{r}{$\sigma$} &
  \multicolumn{1}{r}{[X/H]} &
  \multicolumn{1}{r}{[X/Fe]} &
  \multicolumn{1}{r}{NLTE cor} &
  \multicolumn{1}{l}{Ref. NLTE}\\
\hline
C  & 6  & 0 &     & $   8.50$  & $   4.02$  & $   0.15$  & $  -4.48$  & $  -0.96$ &  & \\
Na & 11 & 0 &   2 & $   6.30$  & $   3.24$  & $   0.20$  & $  -3.06$  & $   0.47$ & --0.35 & \citet{alexeeva2014} \\
Mg & 12 & 0 &   5 & $   7.54$  & $   4.57$  & $   0.13$  & $  -2.97$  & $   0.55$ & 0.24 & \citet{bergemann2017}\\
Al & 13 & 0 &   1 & $   6.47$  & $   2.25$  & $   0.15$  & $  -4.22$  & $  -0.69$ & $\sim 0.4$ & \citet{andrievsky2008}\\
Si & 14 & 0 &   1 & $   7.52$  & $   4.14$  & $   0.15$  & $  -3.38$  & $   0.14$ & 0.03 & \citet{bergemann2013}\\
S  & 16 & 0 &   1 & $   7.16$  & $   4.62$  & $   0.15$  & $  -2.54$  & $   0.98$ & --0.20 & \citet{takeda2005}\\
K  & 19 & 0 &   1 & $   5.12$  & $   2.12$  & $   0.15$  & $  -3.00$  & $   0.52$ & --0.20 & \citet{andrievsky2010}\\
Ca & 20 & 0 &  15 & $   6.33$  & $   3.22$  & $   0.08$  & $  -3.11$  & $   0.41$ & 0.20 & \citet{mashonkina2017}\\
Sc & 21 & 1 &   3 & $   3.10$  & $  -0.21$  & $   0.01$  & $  -3.31$  & $  -0.05$ \\
Ti & 22 & 0 &   9 & $   4.90$  & $   1.68$  & $   0.07$  & $  -3.22$  & $   0.30$ \\
Ti & 22 & 1 &  30 & $   4.90$  & $   1.93$  & $   0.09$  & $  -2.97$  & $   0.29$ & 0.03 & \citet{sitnova2016}\\
V  & 23 & 0 &   3 & $   4.00$  & $   0.23$  & $   0.16$  & $  -3.77$  & $  -0.24$ \\
Cr & 24 & 0 &   8 & $   5.64$  & $   1.95$  & $   0.06$  & $  -3.69$  & $  -0.17$ & 0.56 & \citet{bergemanncr2010}\\
Mn & 25 & 0 &   6 & $   5.37$  & $   1.61$  & $   0.08$  & $  -3.76$  & $  -0.23$ & 0.59 & \citet{bergemannmn2008}\\
Fe & 26 & 0 & 131 & $   7.52$  & $   4.00$  & $   0.09$  & $  -3.52$  & $   0.00$ & 0.21 & \citet{mashonkina2011}\\
Fe & 26 & 1 &  13 & $   7.52$  & $   4.26$  & $   0.09$  & $  -3.26$  & $   0.00$ \\
Co & 27 & 0 &   7 & $   4.92$  & $   1.45$  & $   0.09$  & $  -3.47$  & $   0.05$ & 0.84 & \citet{bergemannco2010}\\
Ni & 28 & 0 &   6 & $   6.23$  & $   2.78$  & $   0.09$  & $  -3.45$  & $   0.08$ \\
Zn & 30 & 0 &   2 & $   4.62$  & $   1.54$  & $   0.02$  & $  -3.08$  & $   0.44$ & 0.15 & \citet{sitnova2022}\\
Sr & 38 & 1 &   2 & $   2.92$  & $   0.57$  & $   0.03$  & $  -2.35$  & $   0.91$ & 0.04 & \citet{mashonkina2022} \\
Y  & 39 & 1 &   8 & $   2.21$  & $  -0.44$  & $   0.09$  & $  -2.65$  & $   0.60$ \\
Zr & 40 & 1 &   7 & $   2.62$  & $   0.21$  & $   0.11$  & $  -2.41$  & $   0.85$ \\
Ba & 56 & 1 &   4 & $   2.17$  & $   0.21$  & $   0.25$  & $  -1.96$  & $   1.30$ & --0.02 & \citet{mashonkina2019}\\
La & 57 & 1 &  30 & $   1.14$  & $  -0.77$  & $   0.10$  & $  -1.91$  & $   1.35$ \\
Ce & 58 & 1 &  72 & $   1.61$  & $  -0.51$  & $   0.09$  & $  -2.12$  & $   1.14$ \\
Pr & 59 & 1 &  11 & $   0.76$  & $  -1.15$  & $   0.09$  & $  -1.91$  & $   1.34$ \\
Nd & 60 & 1 & 112 & $   1.45$  & $  -0.36$  & $   0.10$  & $  -1.81$  & $   1.45$ \\
Sm & 62 & 1 &  58 & $   1.00$  & $  -0.65$  & $   0.07$  & $  -1.65$  & $   1.61$ \\
Eu & 63 & 1 &   5 & $   0.52$  & $  -0.90$  & $   0.07$  & $  -1.42$  & $   1.84$ & 0.23 & \citet{mashonkina2000} \\
Gd & 64 & 1 &  23 & $   1.11$  & $  -0.53$  & $   0.06$  & $  -1.64$  & $   1.62$ \\
Dy & 66 & 1 &  14 & $   1.13$  & $  -0.36$  & $   0.12$  & $  -1.49$  & $   1.76$ \\
Er & 68 & 1 &   6 & $   0.96$  & $  -0.64$  & $   0.12$  & $  -1.61$  & $   1.65$ \\
Lu & 71 & 1 &   2 & $   0.12$  & $  -1.12$  & $   0.02$  & $  -1.24$  & $   2.02$ \\
Hf & 72 & 1 &   1 & $   0.87$  & $  -0.98$  & $   0.15$  & $  -1.85$  & $   1.40$ \\
Th & 90 & 1 &   2 & $   0.08$  & $  -1.33$  & $   0.11$  & $  -1.41$  & $   1.85$ & 0.15 & \citet{mashonkina2012}\\
U  & 92 & 1 &   1 & $ -0.52 $  & $ <-1.8 $  &            & $ < -1.28$ & $ <2.24 $ \\
\hline\end{tabular}
\end{table*}

\section{Discussion}

The star \ymp\ is: (1) rich in n-capture elements, (2) rich in the actinide element Th, and (3) poor in carbon.
Both these two chemical situations are particularly rare.
In the EMP regime, just a few stars, the large majority of them more evolved than \ymp, are known with such a low [C/Fe]. 
It may be tempting to make the hypothesis that the r-II nature and the low [C/Fe] are linked. 
In fact, some C-poor EMP stars have been analysed by \citet{hansen2018}; they are all more metal-rich, mainly more evolved, and the majority are not so low in [C/Fe] as \ymp.
The very MP star 2MASS\,J17045729+3720576 \citep{bandyopadhyay2024}, an r-II according to \citet{holmbeck2020} and an r-I according to \citet{christlieb2004}, 
is even more C-poor than \ymp.
These facts may suggest that poor C content is somehow related to the production of r elements.
Indeed the spectra of C-poor stars are easily analysed: no or very few CH and CN features contaminate the observed spectra, making easy to derive abundances of heavy elements.
We then expect a bias against CEMP r-I or r-II stars, but we do not expect such a bias against r-I and r-II for C-normal stars.

\begin{figure}[hb]
    \centering
   \resizebox{0.48\textwidth}{!}{\includegraphics{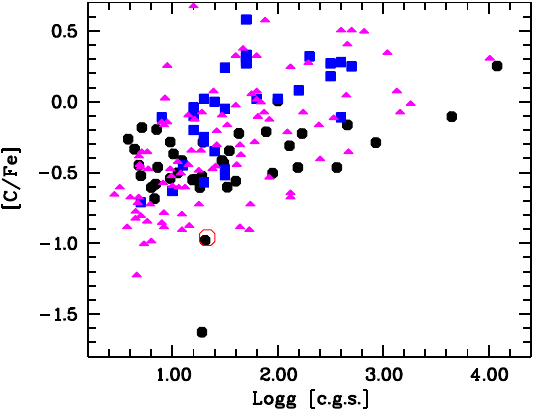}}
   \caption{[C/Fe] ratio of \ymp\ (open red) compared with a sample of MP stars \citep[filled black circles][]{rvs3}, a sample of EMP stars \citep[filled blue squares][]{cayrel2004},
 and the sample from \citet[][filled pink triangles]{hansen2018}.
    }
    \label{fig:clogg}
\end{figure}

In Fig.\,\ref{fig:clogg} the [C/Fe] ratio versus the surface gravity for \ymp\ is compared to a more metal-rich ($ \rm\langle[Fe/H]\rangle =-1.76$) stellar sample selected for their high radial-velocity by \citet{rvs3}. 
All but one star in this sample are likely mixed, so that $\rm [C/Fe]<0$ is expected. Two stars stand out for having a very low [C/Fe] ratio: RVS740 with $\rm [Fe/H]=-2.86$ and $\rm [C/Fe]=-0.98$; RVS718 with $\rm [Fe/H]=-1.65$ and $\rm [C/Fe]=-1.63$.
The stars in the EMP reference sample of \citet{cayrel2004} have metallicities similar to \ymp, but no star is so low in [C/Fe] \citep{spite2005}. 
The stars in \citet{hansen2018} are all more metal-rich than \ymp\ ($\rm -3.12 < [Fe/H]< -0.7$), but some of them are in fact almost as low as \ymp\ in [C/Fe].
One can always invoke extra mixing, above what is normally observed above the red giant branch (RGB) bump, to explain this extremely low C abundance, yet this hypothesis is not
supported by anything else, observationally or theoretically.

\begin{figure}[ht]
    \centering
   \resizebox{0.48\textwidth}{!}{\includegraphics{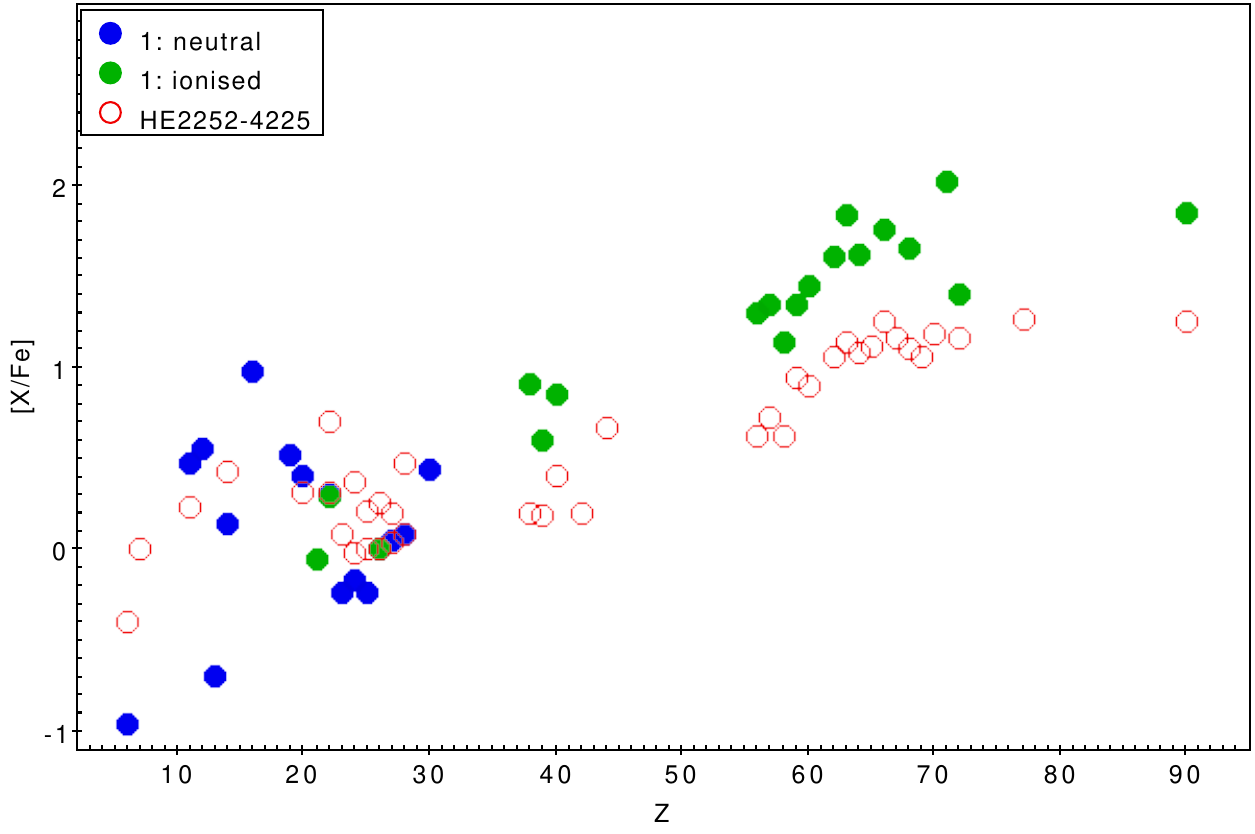}}
   \caption{Abundances of \ymp\ compared to HE\,2252--4225 \citep{mashonkina2014x}.
    }
    \label{fig:comphe2252}
\end{figure}

\begin{figure}[ht]
    \centering
   \resizebox{0.48\textwidth}{!}{\includegraphics{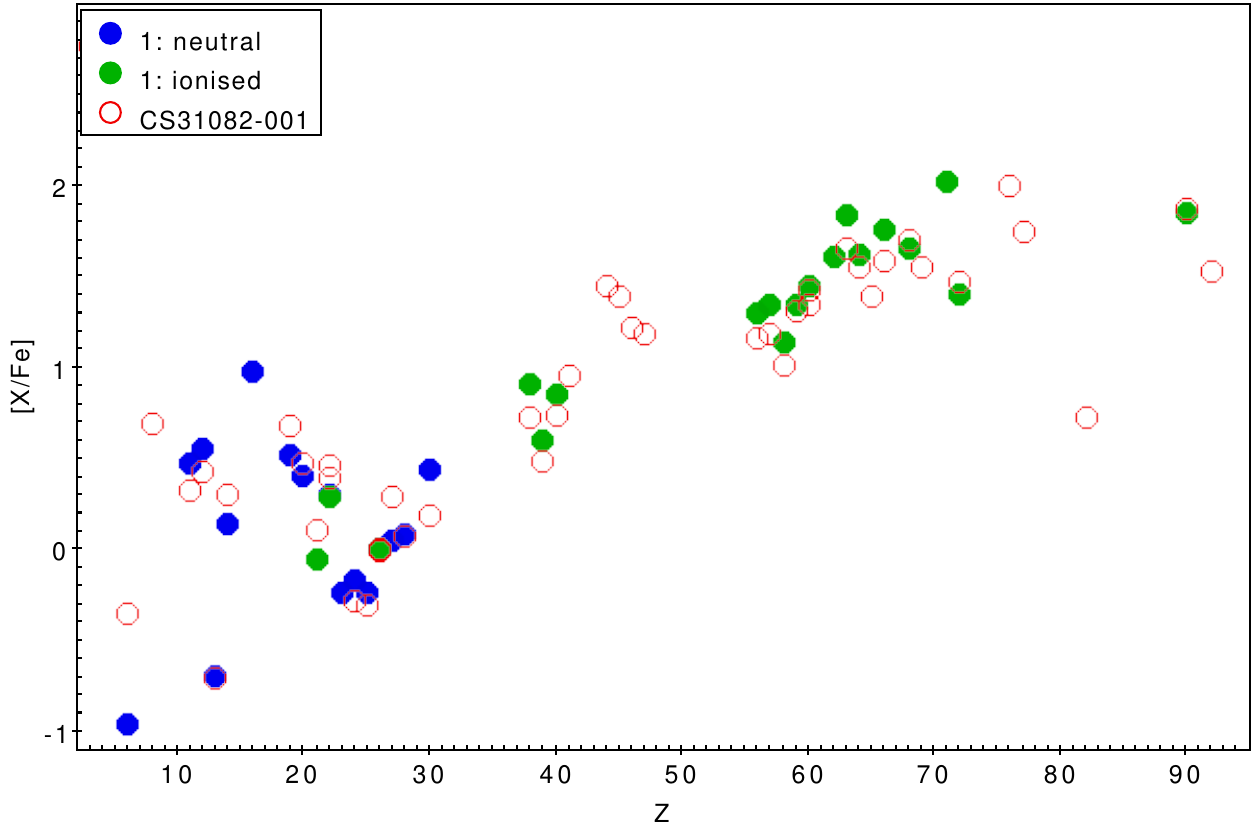}}
   \caption{Abundances of \ymp\ compared to CS\,31082--001 \citep{hill2002}.
    }
    \label{fig:compcs31082}
\end{figure}

Known r-II stars 
show some diversity in their abundance patterns; this can be seen in Fig.\,\ref{fig:comphe2252}, where the pattern
of HE\,2252--4225 \citep[actinide boost according to the definition of][]{mashonkina2014x} is compared to that of \ymp.
The [X/Fe] ratios for the elements Ca to Zn are by and large comparable. The two patterns for heavy elements ([X/Fe]) are offset by about  0.5\,dex, \ymp\ being higher in [X/Fe]. The elements C to K again also show a surprising diversity.
In Fig.\ref{fig:compcs31082}, we compare the [X/Fe] ratios in \ymp\ with those of the prototype actinide
boost star CS\,31082--001 \citep{cayrel2001,hill2002}.
This plot shows some striking similarities, namely among the n-capture elements, 
in particular Th, and also some clear differences among the light elements,
in particular carbon, which is underabundant by almost 1\,dex with respect to iron
in \ymp, while in CS\,31082--001 it is depleted as expected by the evolutionary state of this star.
From Fig.\,\ref{fig:comphe2252} and Fig.\ref{fig:compcs31082}, one concludes that the chemical pattern in heavy elements for two r-II stars defined as actinide boost (HE\,2252--4225 and CS\,31082--001) are quite different. 
This diversity is a strong motivation for having as many details as possible on the chemical pattern of these r-rich stars, to be able to solve the quest on where and when the n-capture (and the actinide) elements are formed.

Not many Galactic EMP  stars \citep[about 20 the SAGA database][]{SAGA} have a A(Th) determination, and only a subsample  
of them has $\rm [Th/Fe]>1$; they are mainly centred at a higher metallicity than \ymp\ ($\rm [Fe/H]\sim -3$).
Of these, only three stars have a metallicity around --3.5: SMSS\,J200322.54--114203.3 \citep{yong2021} with $\rm [Fe/H]=-3.57$ and $\rm [Th/Fe]=2.2$;
SPLUS\,J142445.34--254247.1 \citep{placco2023} with $\rm [Fe/H]=-3.41$ and $\rm [Th/Fe]=2.12$;
CS\,30315--029 \citep{siqueiramello2014} whose [X/Fe] ratios for the n-capture elements are much lower than for \ymp.

The chemical pattern of \ymp\ is similar to what has been derived for other stars.
In fact, the abundance ratios [X/Fe] are similar to: SMSS\,J200322.54--114203.3 \citet{yong2021},
CS\,29497--005 \citep{christlieb2004},
CS\,31081--001 \citep{hill2002},
HE\,1219--0312 \citep{roederer2009},
RAVE\,J203843.2--002333 \citep{placco2017},
and CS\,29529--0089 \citep{dasilva2025}.
None of these stars is poor in carbon, though.

As the number of known actinide boost\footnote{The number of actinide boost stars depends on
the definition adopted for actinide boost.} and r-II stars increases, the diversity among them also increases. In the case of \ymp\ , it is the
low carbon abundance that is at odds with the majority of the observed stars. 
Also the difference in the pattern of heavy elements  between \ymp\ and other r-II stars is evident.
 
\begin{figure}[ht]
    \centering
   \resizebox{0.48\textwidth}{!}{\includegraphics{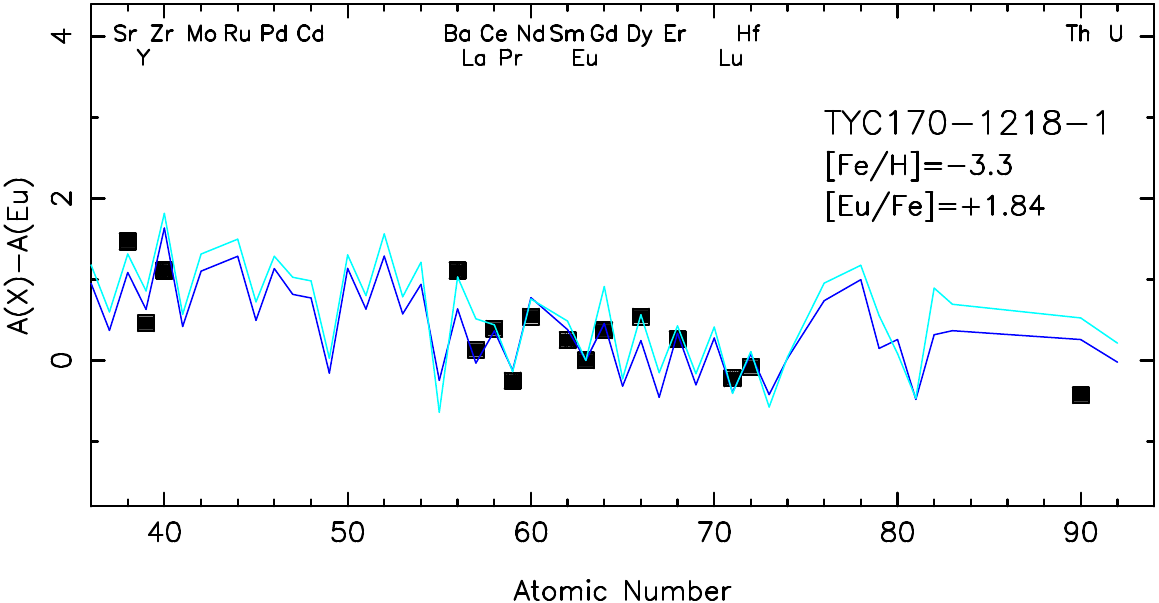}}
   \resizebox{0.48\textwidth}{!}{\includegraphics{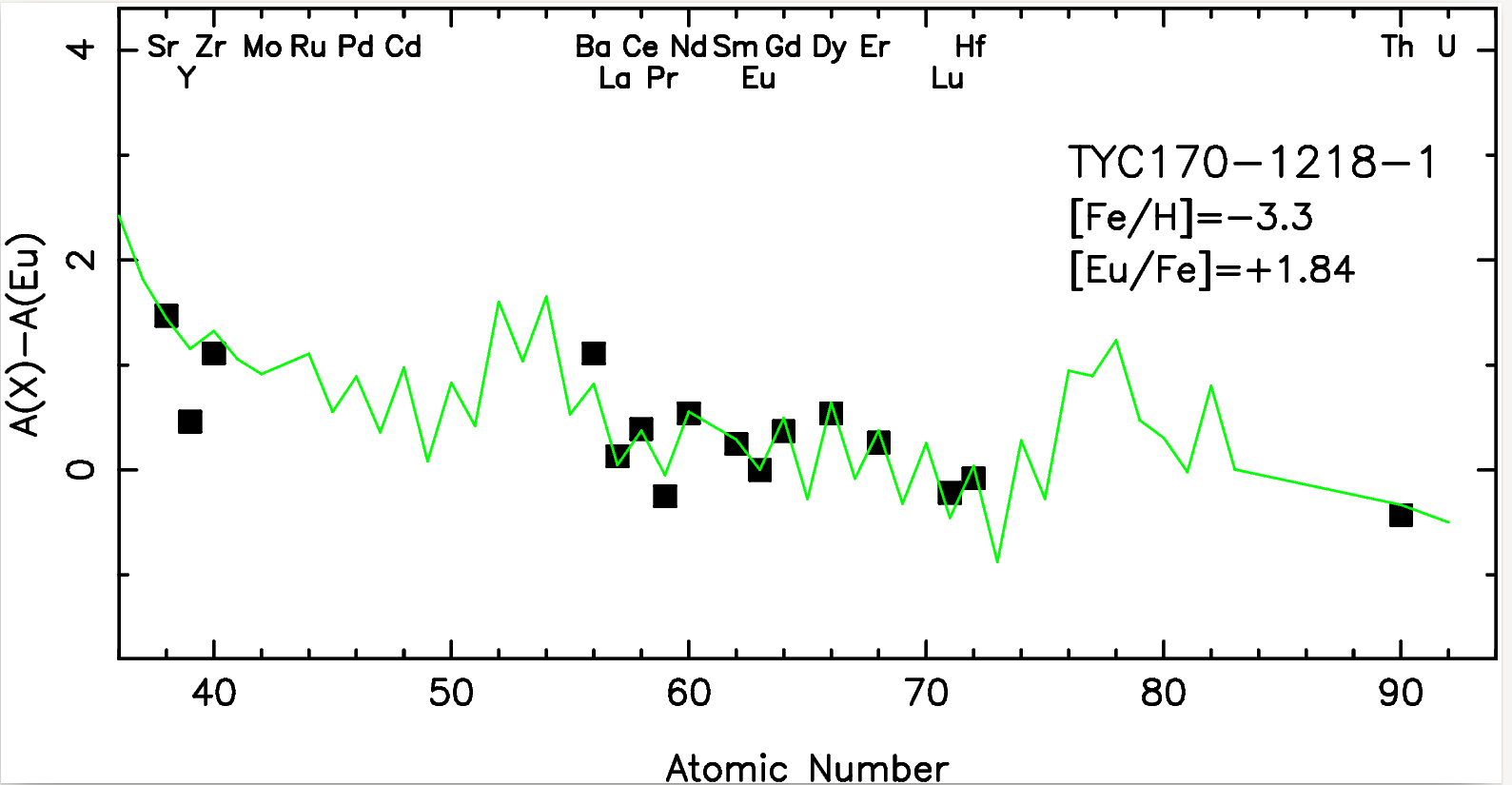}}
   \caption{A(X)--A(Eu) in \ymp\ compared to: (upper panel) the model of Wanajo \citep{siqueiramello2013} computed for two different temperatures (solid blue cold model and solid cyan hot model); (lower panel) the distribution of the solar r process \citep{thielemann2026}.
    }
    \label{fig:compwanajo}
\end{figure}

In the top panel of Fig.\,\ref{fig:compwanajo}, the abundances of \ymp\ are compared to the hot and cold models \citep{wanajo2007,siqueiramello2013}. 
The difference in the radioactive Th is due to its decay since the formation of the star. 
In the lower panel, we show the comparison with the solar r-process abundances from \citet{thielemann2026}.
The fact that the solar Th/Eu ratio (almost identical to the solar r-process ratio, since both
elements are essentially formed by the r process) is so close to the value found in \ymp\  
is remarkable, considering the age difference between the two stars. This implies that at the
origin the Th content in \ymp\ was much higher.
For 23 r-II stars \citep[according the definition by][]{christlieb2004} from the literature and \ymp, the $\rm A(Th)-A(Eu)$
varies from $-0.84$ to 0.04 ($\rm A(Th)-A(Eu)=-0.43$ in \ymp) 
with $\rm\langle\left(A(Th)-A(Eu)\right)\rangle =-0.48\pm 0.21$, to be compared to the solar value of $-0.44$.
These very or extremely MP stars are supposed to be much older ($\sim 6-9$\,Gyr) than the Sun. Their Th has decayed more than for the Sun.
It is reasonable to think that for the stars with a slightly sub-solar or an over-solar $\rm A(Th)-A(Eu)$, Th have been produced
by an extra channel with respect to the r process.

 \section{Conclusions}
 
The three main results of our investigation are: {\em i)} to firmly establish that star \ymp\ is a 
neutron-capture rich and Th rich star; 
{\em ii)} that it is very poor in carbon, more than what can be justified by a depletion due to standard mixing and more than what is normally
observed above the RGB bump; 
{\em iii)} that it belongs to the Sequoia accretion event, and thus was formed in an external galaxy and not in the Milky Way.
The extremely low carbon abundance could be explained
by some extra mixing; however, it is interesting to consider the possibility that
the star was initially formed with a low carbon abundance, which was further depleted during
the evolution of the star.
 
Being able to determine the U abundances in all r-II stars would be extremely important for our understanding of the actinide production mechanism.
Stars such as  HE\,2252--4225 are too faint for the currently available telescopes,
but U measurement may become achievable once the ANDES spectrograph on the ELT is available \citep{andes}.
However, \ymp\ is bright enough that further eight hours of integration
with UVES or another high-resolution spectrograph on an 8\,m class telescope should allow its U abundance to be determined, provided its [U/Fe] is similar to that of
CS\,31082--001. This is a reasonable expectation, since their [Th/Fe] are almost the same.

\begin{acknowledgements}
The authors wish to thank R. Lallement and Y. Lebreton for their help and useful discussion.
We thank the anonymous referee for the help in improving the manuscript.
This work has made use of data from the European Space Agency (ESA) mission
{\it Gaia} (\url{https://www.cosmos.esa.int/gaia}), processed by the {\it Gaia}
Data Processing and Analysis Consortium (DPAC,
\url{https://www.cosmos.esa.int/web/gaia/dpac/consortium}). Funding for the DPAC
has been provided by national institutions, in particular the institutions
participating in the {\it Gaia} Multilateral Agreement.
This research has made use of the SIMBAD database, operated at CDS, Strasbourg, France. LM gratefully acknowledges support from ANID FONDECYT Regular Project n. 1251809.
\end{acknowledgements}




%


  \bibliographystyle{aa} 
 \bibliography{biblio} 

%



\end{document}